\documentclass[journal,final,twoside]{IEEEtran}
\usepackage{cite}

\ifCLASSINFOpdf
   \usepackage[pdftex]{graphicx}
   \graphicspath{{./figures/}{./pdf/}{./jpeg/}}
   \DeclareGraphicsExtensions{.pdf,.jpeg,.png}
\else
   \usepackage[dvips]{graphicx}
   \graphicspath{{./eps/}}
   \DeclareGraphicsExtensions{.eps}
\fi
\usepackage{algorithm}
\usepackage{algpseudocode}

\usepackage[cmex10]{amsmath}
\usepackage{amssymb}

\usepackage[caption=false,font=footnotesize]{subfig}
\usepackage{stfloats}

\usepackage{multirow}

\usepackage{footnote}

\usepackage{threeparttable}

\renewcommand{\vec}[1]{\mathbf{#1}}

\begin{document}
%
\title{An Arithmetic Coding Scheme for Blocked-based Compressive Sensing of Images}
%
%
%

\author{Min Gao
\thanks{Min Gao is with the Department of Computer Science and Technology, Harbin Institute of Technology, China. E-mail address: mgaohitcs@outlook.com and mgao@hit.edu.cn.}
}

%
%

\markboth{}%
{An Arithmetic Coding Scheme for Blocked-based Compressive Sensing of Images}
%



\maketitle

\begin{abstract}
Differential pulse-code modulation (DPCM) is recently coupled with uniform scalar quantization (SQ) to improve the rate-distortion (RD) performance for the block-based quantized compressive sensing (CS) of images. In this framework, for each block's CS measurements, a prediction is generated based on the reconstructed CS measurements of the previous blocks and subtracted from measurements of the current block in the measurement domain. The resulting residual is then quantized by uniform SQ to generate the quantization index. However, the entropy coding is still required to remove the statistical redundancies between the quantization indices and generate the bitstream. Thus, in this paper, we proposed an arithmetic coding scheme for the quantization index within DPCM-plus-SQ framework by analyzing their statistics. Experimental results demonstrate that further RD performance can be achieved compared to original DPCM-plus-SQ scheme and transform coefficient coding in CABAC.
\end{abstract}



%
\IEEEpeerreviewmaketitle

\section{Introduction}
\IEEEPARstart{C}{ompressive} (CS) [1] is an emerging framework, which allows the measurements of a sparse signal to be sampled below Nyquist rate and guarantees exact recovery of the signal from these measurements with high probability. Different from the traditional point-by-point sampling, the CS measurement process is assumed to be accomplished within hardware of sensing device via a linear projection into a lower-dimensional subspace chosen at random. Consequently, the CS measurement process is regarded as a joint sampling and compression approach.  However, the CS measurement process is not yet a real compression technique in the strict information theoretic sense [2], since a real compression technique is generally considered to convert the input data into a compressed bitstream. Indeed, the quantization and entropy coding techniques are required to generate the compressed bitstream from the CS measurements by removing the redundancies.

There have been some solutions to incorporate the quantization into CS framework. The straightforward solution is to directly apply the scalar quantization (SQ) to the CS measurements sampled by sensing device. Nonetheless, the rate-distortion (RD) performance of this solution has been demonstrated inefficient [2, 3], since the characteristics of CS and signal itself are not exploited. As a result, various techniques have been recently proposed to improve the RD performance of the quantized CS, mainly depending on the quantization optimization process [4], the reconstruction process [5, 6] or both [7, 8].
\newline
\vspace{2pt}
\hspace{6pt}In contrast to the works mentioned above, a framework of quantization via simple uniform SQ coupled with differential pulse code modulation (DPCM) of the CS measurements was recently proposed for block-based CS (BCS) in [2]. In this framework, the reconstructed measurements of the previous block is used as a prediction, and subtracted from the measurements of the current block in measurement domain. Then the resulting residual is quantized with uniform SQ to generate the quantization index. According to the experimental results provided in [2], this simple DPCM-plus-SQ approach can provide competitive RD performance compared with the approaches in [6, 8]. To further improve the RD performance, some advanced prediction techniques were proposed in [9, 10] by utilizing the spatial correlation of images, in which the optimal prediction for each block¡¯s CS measurements is selected from a set of multiple prediction candidates.
\newline
\vspace{2pt}
\hspace{6pt}After obtaining the quantization index, the entropy coding is still required to remove the statistical redundancies and generate the bitstream. Although there have been several existing entropy coding schemes for image/video coding such as EBCOT [11] for JPEG2000 and CABAC [12] for H.264/AVC, they are not suitable to compress the quantization index of CS measurements due to the statistical difference between CS measurements and the transform coefficients in image/video coding. Thus, it is desirable to develop an entropy coding scheme for the quantization index of CS measurements.
\newline
\vspace{2pt}
\hspace{6pt}Inspired by the arithmetic coding scheme for the transform coefficients in CABAC, we propose an arithmetic coding scheme in this paper for the quantization index of CS measurements by analyzing their statistics. In the proposed arithmetic coding scheme, the quantization index of CS measurements is represented by three syntax elements, namely \textit{significant\_map}, \textit{abs\_coeff\_level\_minus1} and \textit{sign\_flag}, which denote the significance, absolute value and sign for a quantization index, respectively. The designed syntax elements are finally coded with binary arithmetic coding engine M-coder [12] to remove their statistical redundancy. To the best of our knowledge, this is the first time that arithmetic coding scheme is designed for the framework of BCS of images. Experimental results demonstrate the efficiency of the proposed arithmetic coding within the BCS framework of images.
\newline
\vspace{2pt}
\hspace{6pt}This paper is organized as follows. Section II briefly reviews CS, BCS and DPCM-plus-SQ framework in [2]. In section III, the details of the proposed arithmetic coding scheme are provided. Section IV presents the experimental results. Section V concludes this paper.
\begin{figure*}[!htbp]
    \includegraphics[width=\textwidth]{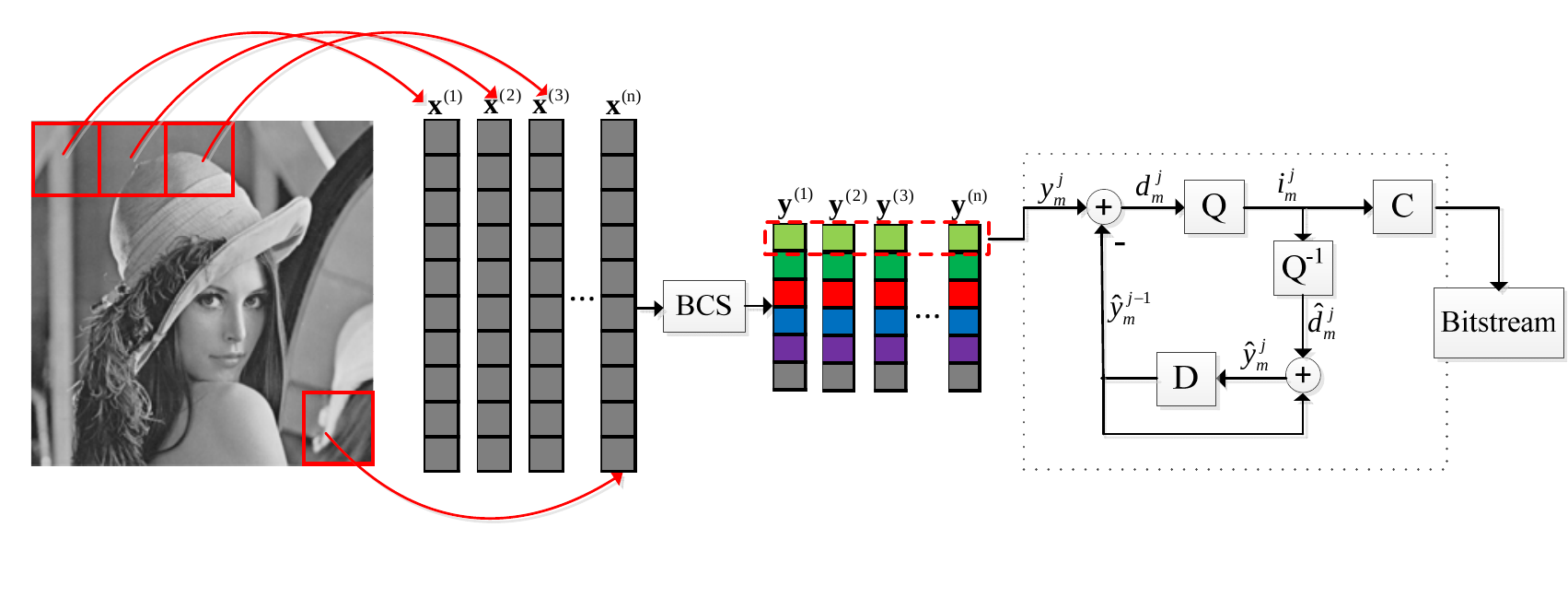}
 	\caption{Framework of DPCM-plus-SQ of block-based compressive sensing (BCS) in [2]. BCS is implemented based on any CS-based acquisition. Q: uniform SQ; C: entropy coded module; D is DPCM prediction module.}
 	\label{Fig:DPCMSQ}
 \end{figure*}

\section{Background}
\label{sec:Background}
\subsection{Block-based compressive sensing of images}
Compressive sensing (CS) is an emerging mathematic paradigm, in which signals is acquired via linear projection into a dimension much lower than that of the original signal and the exact recovery of the signal can be guaranteed if the signal is sparse in some domain. More specifically, suppose that we have a real-valued signal $ \vec{x} \in \mathbb{R} ^N $ and its measurements $ \vec{y} \in \mathbb{R} ^M $ are obtained by the linear projection $ \vec{y} = \vec{\Phi} \vec{x}$ with sampling rate or $subrate$ $S = \frac{M}{N}$. Here, $\vec{\Phi}$ is a $M \times N$ measurement matrix such that $M$ is much smaller than $N$. CS theory allows the exact recovery of $\vec{x}$ from $\vec{y}$, if $\vec{x}$ is sparse in some domain.

In order to avoid the large storage cost of measurement matrix in CS of 2D images, the block-based CS (BCS) was proposed in [13], in which the sampling of an image is driven by the random matrix applied on a block-by-block basis. That is, an image $\vec{x}$ is first partitioned into n non-overlapped $B \times B$ blocks and each block is denoted by $\vec{x}^{\left(j\right)} \in B^2$ with $j = 1, 2, ..., n$ in raster scan fashion. Then the corresponding measurement vector $\vec{y}^{\left(j\right)}$ for $\vec{x}^{\left(j\right)}$ is obtained by
\begin{IEEEeqnarray}{rCl}
	\label{equ1}
	\vec{y}^{\left(j\right)} = \left[ y_{1}^{j},..., y_{m}^{j},...,y_{M_B}^{j}\right] = \vec{\Phi}_B \vec{x}^{\left(j\right)},
\end{IEEEeqnarray}
where $\vec{y}^{\left(j\right)} \in \mathbb{R} ^{M_{B}} $ and $\vec{\Phi}_B$ is a $M_B \times B^2$ measurement matrix such that the $subrate$ for the block is $\frac{M_B}{B^2}$. It is straightforward to see that $\vec{\Phi_{B}}$ applied to an image on block-by-block basis is equivalent to the whole image measurement matrix $\vec{\Phi}$ with a constrained structure. That is $\vec{\Phi}$ can be written as block diagonal with $\vec{\Phi_{B}}$ along the diagonal [2].

\subsection{DPCM-plus-SQ framework for CS compression}
To improve the RD performance of quantized CS measurement, the DPCM-plus-SQ framework was proposed in [2] for BCS compression. As shown in Fig. \ref{Fig:DPCMSQ}, an input image $\vec{x}$ is first divided into $n$ non-overlapped $B \times B$ blocks, denoted by $\vec{x}^{\left(j\right)}$ with $j = 1, 2,..., n$ in raster scan fashion. Then the CS measurements for all blocks are acquired with Eq. (1). Next, for block $\vec{x}^{\left(j\right)}$ of the image, the $m$th component $y_{m}^{j}$ in the measurement vector $\vec{y}^{\left(j\right)}$ is predicted by the corresponding vector component $\hat{y}_m^{j - 1}$ in the reconstructed measurement vector $\hat{\vec{y}}^{\left(j - 1\right)}$ of the previously processed block $\vec{x}^{\left(j - 1\right)}$. Then, the resulting residual $d_m^j = y_m^j - \hat{y}_m^{j - 1}$ is scalar quantized to generate the quantization index $i_m^j = Q\left[d_m^j\right]$, which is finally entropy coded. The prediction feedback loop required at the encoder consists of the de-quantization of $i_m^j$ producing the quantized residual $\hat{d}_m^j$ such that $\hat{y}_m^j = \hat{d}_m^j + \hat{y}_m^{j - 1}$. So the prediction can be implemented on the block-by-block basis with one block delay buffer.

Actually, the quantization indices are not entropy coded in [2] and the zero order entropy of the quantization indices is used as an estimate of the actual bitrate that would be produced by a real entropy coder, which is calculated as follows:

\begin{IEEEeqnarray}{rCl}
    \label{equ2}
    E = \sum_{y}p\left(y\right) \log_2{\left[p\left(y\right)\right]}
\end{IEEEeqnarray}
where $p\left(y\right)$ is the probability of quantization index equal to $y$.

\section{Arithmetic coding for blocked-based compressive sensing}
\label{sec:ProposedWork}
\begin{figure}
    \includegraphics[width=0.45\textwidth]{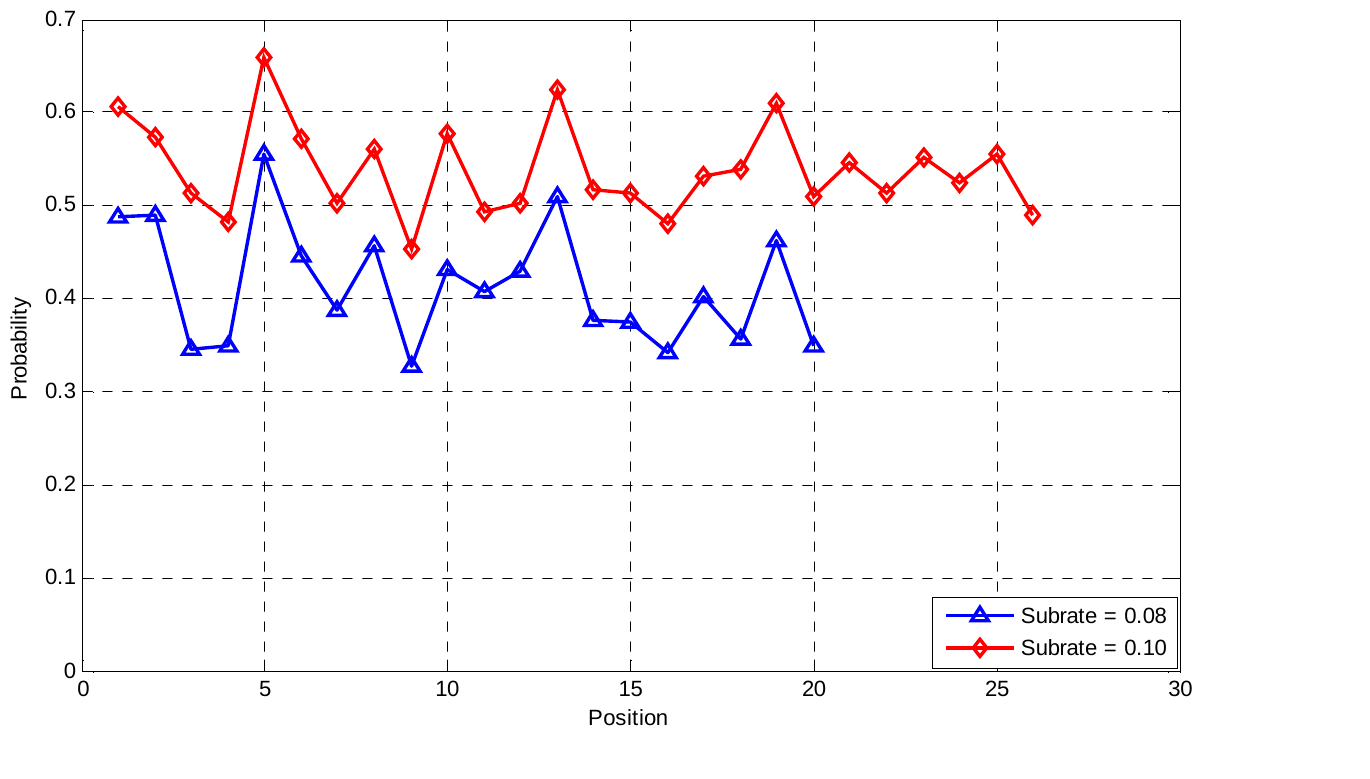}
 	\caption{Occurrence probability of the significant component for each position within a quantization index vector.}
 	\label{Fig:SIGPDF}	
\end{figure}

In the coding of quantization index vector $\vec{i}^j = [i_1^j,...,i_m^j,...,i_{M_B}^j]^T$, we find that the occurrence probability of the significant components is independent of their position within the quantization index vector, as shown in Fig. \ref{Fig:SIGPDF}, where the data is collected from BCS of Lena at $subrate$ 0.08 and 0.10. From the statistical information on the quantization index vectors, we also find that the last component in the quantization index vector is likely to be significant. In other words, the significant component is randomly distributed throughout the quantization index vector. So, we use a syntax element namely \textit{significant\_map} to indicate the significance of each component within the quantization index vector. For each significant component, the syntax elements \textit{abs\_coeff\_level\_minus1} and \textit{sign\_flag} are signaled to indicate its absolute magnitude and sign, respectively. Table. \ref{table:Table1} lists the syntax elements in the proposed arithmetic coding scheme.

\begin{table} [h]
	\renewcommand{\arraystretch}{1.3}
	\caption{Syntax elements used in the proposed arithmetic coding scheme}
	\label{table:Table1}
	\centering
	\resizebox{0.5\textwidth}{!}{%
		\begin{tabular}{|c||c|}
			\hline
			\bfseries \shortstack{\\Syntax Elements} & \bfseries \shortstack{\\Defination} \\
			\hline\hline
			\shortstack{\\\textit{significant\_map}\\~} & \shortstack{Indicate the significance for each component \\ within a quantization index vector.}\\
			\hline
            \shortstack{\\\textit{abs\_coeff\_level\_minus1}\\~} & \shortstack{Indicate the absolute value of significant \\ component minus one within a quantization \\ index vector.}\\
			\hline
            \shortstack{\\\textit{sign\_flag}\\~} & \shortstack{Indicate the sign of significant component within a \\ quantization index vector.}\\
			\hline
		\end{tabular}
	}
\end{table}

Based on the designed syntax elements, the coding of a quantization index vector is summarized as follows. First, the \textit{significant\_map} is first transmitted for each component in the quantization index vector to indicate the location of the significant components. If the \textit{significant\_map} for component $i_m^j$ is equal to one, it means that $i_m^j$ is significant component. Then, the \textit{abs\_coeff\_leve\_minus1} and \textit{sign\_flag} are signaled for each significant component to indicate its magnitude and sign, respectively.

To efficiently code these syntax elements, the binary arithmetic coding engine M-coder in CABAC [12] is adopted in the proposed arithmetic coding scheme. M-coder is a multiplication-free binary arithmetic coder and the probability update module in M-coder can adaptively update the probability according to the previously coded symbols, which can capture the local statistical features of a binary source. More specifically, the probability of a symbol to be coded is presented by $\left(p_{LPS}, V_{MPS}\right)$ in M-coder, where $p_{LPS}$ denotes the probability of least probable symbol (LPS) and $V_{MPS}$ denotes the value of the most probable symbol (MPS). The range of $p_{LPS}$ is projected into a set of representative probabilities $S_{p}=\{p_0, p_1,...,p_{63}\}$ and each representative probability $p_{\sigma}$, $0 \leq \sigma \leq 63$ is implicitly represented by its index $\sigma$. As a result of this design, only two parameters $(\sigma, V_{MPS})$ are required to be determined when using M-coder to code a given binary source.

The syntax element \textit{abs\_coeff\_level\_minus1} is mapped into a string of bins (binary symbols) using UEG0 binarization scheme [12], since it is non-binary valued symbol. UEG0 is specified by the cutoff value $S = 14$ for the truncated unary prefix part and the order $k = 0$ for the Exp-Golomb suffix part.

Since the CS measurements are generated via the linear projection onto random basis, the relationship between different components within a quantization index vector is also random. In other words, the probability distribution of the components at different positions is independent of each other. Consequently, for coding \textit{significant\_map} and \textit{abs\_coeff\_level\_minus1}, only one context model is used for each syntax element. For \textit{sign\_flag}, the equal probability (probability equal to 0.5) is used.

The differences between the proposed arithmetic coding scheme and the transform coefficient coding in CABAC mainly reflect the following aspects. First, the transform coefficient coding in CABAC uses the syntax element \textit{last\_significant\_map} to indicate the position of the last significant coefficient for a transform block, while the syntax element \textit{last\_significant\_map} was removed in the proposed method. This is mainly because it is meaningless to encode \textit{last\_significant\_map} in coding of quantization index within BCS framework, since the last significant component is likely to be at the end of the quantization index vector. Second, there are different context models for \textit{significant\_map} and \textit{abs\_coeff\_level\_minus1} in the transform coefficient coding in CABAC and the context model selection is dependent on the coefficient position and the previously coded transform coefficients, while there is only one context model in the proposed arithmetic coding scheme for \textit{significant\_map} and \textit{abs\_coeff\_level\_minus1}. This is because the context modeling in CABAC will cause the context dilution problem within BCS framework, since the occurrence probability of the significant components is independent of their position and the previously coded quantization indexes do not have any correlation with the current one.

\section{Experimental Results}
\label{sec:Experimental Results}
In our experiments, four 512x512 grayscale images are used to verify the performance of the proposed arithmetic coding scheme, which include $Lenna$, $Barbara$, $Goldhill$ and $Peppers$. The block size for BCS is set to be $16 \times 16$, and the measurement matrix $\vec{\Phi}_B$ is an orthogonal random Gaussian matrix. The setup of the combination of quantizer step-size and $subrate$ is the same as that in [2]

We implement the proposed arithmetic coding scheme within the DPCM-plus-SQ framework in [2], which is referred to as AC-DPCM-SQ in the following description. For the comparison, we also implement the transform coefficient coding in CABAC within the DPCM-plus-SQ framework, which is referred to as CABAC-DPCM-SQ. The quantization index vectors in AC-DPCM-SQ and CABAC-DPCM-SQ are signaled into the arithmetic coding module to generate the bitstream and calculate the actual bitrate, while the quantization index vectors in the original DPCM-plus-SQ method in [2] (referred to as ORG-DPCM-SQ) are not actually entropy coded and the zero order entropy of the quantization index is used as the estimate of the actual bitrate, which is calculated according to Eq. \ref{equ2}.

Since the arithmetic coding scheme is a lossless coding method, the distortion of a given image in AC-DPCM-SQ and CABAC-DPCM-SQ is the same as that in ORG-DPCM-SQ. Thus Tables \ref{table:Table2} and \ref{table:Table3} gives the bitrate reduction ($BR$) at different $subrates$ by comparing AC-DPCM-SQ with CABAC-DPCM-SQ and ORG-DPCM-SQ, respectively. In Tables \ref{table:Table2} and \ref{table:Table3}, $Meth1$, $Meth2$ and $Meth3$ represent ORG-DPCM-SQ, CABAC-DPCM-SQ and AC-DPCM-SQ, respectively. $BR_{1,3}$ and $BR_{2,3}$ in Tables 2 and 3 are the bitrate reduction achieved by AC-DPCM-SQ over ORG-DPCM-SQ and CABAC-DPCM-SQ, respectively. As shown in Tables \ref{table:Table2} and \ref{table:Table3}, compared with ORG-DPCM-SQ, the bitrate reduction achieved by AC-DPCM-SQ is between 3.66\% and 10.62\% for different images. Compared with CABAC-DPCM-SQ, the bitrate reduction achieved by AC-DPCM-SQ is between 2.37\% and 5.56\% for different images.
\begin{table*} []
	\renewcommand{\arraystretch}{1.3}
	\caption{Bitrate reduction ($BR$) achieved by AC-DPCM-SQ for $Lenna$ and $Barbara$.}
	\label{table:Table2}
	\centering
	\begin{tabular}{|c||c|c|c|c|c||c|c|c|c|c|}
		\hline
		\multirow{2}{*}{\bfseries \shortstack{Subrate\\Index}} & \multicolumn{5}{|c||}{\bfseries \shortstack{$Lenna$}} &  \multicolumn{5}{|c|}{\bfseries\shortstack{$Barbara$}} \\
        \cline{2-11}
        & \bfseries \shortstack{$Meth1$\\$[bpp]$} & \bfseries \shortstack{$Meth2$\\$[bpp]$} & \bfseries \shortstack{$Meth3$\\$[bpp]$} & \bfseries \shortstack{$BR_{1,3}$\\$[\%]$} & \bfseries \shortstack{$BR_{2,3}$\\$[\%]$} & \bfseries \shortstack{$Meth1$\\$[bpp]$} & \bfseries \shortstack{$Meth2$\\$[bpp]$} & \bfseries \shortstack{$Meth3$\\$[bpp]$} & \bfseries \shortstack{$BR_{1,3}$\\$[\%]$} & \bfseries \shortstack{$BR_{2,3}$\\$[\%]$} \\
        \hline\hline
        1 & 0.150 &	0.144 & 0.137 &	\textbf{8.67} &	\textbf{4.86} &	0.159 &	0.156 &	0.152 &	\textbf{4.40} &	\textbf{2.56} \\
        \hline
        2 &	0.258 &	0.248 &	0.234 &	\textbf{9.30} &	\textbf{5.65} &	0.245 &	0.241 &	0.231 &	\textbf{5.71} &	\textbf{4.15} \\
        \hline
        3 & 0.313 &	0.302 &	0.278 &	\textbf{11.18} & \textbf{7.95} &	0.308 &	0.302 &	0.288 &	\textbf{6.49} &	\textbf{4.64} \\
        \hline
        4 &	0.418 &	0.400 &	0.374 &	\textbf{10.53} & \textbf{6.50} &	0.386 &	0.378 &	0.360 &	\textbf{6.74} &	\textbf{4.76} \\
        \hline
        5 &	0.554 &	0.527 &	0.493 &	\textbf{11.01} & \textbf{6.45} &	0.551 &	0.536 &	0.513 &	\textbf{6.90} &	\textbf{4.29} \\
        \hline
        6 &	0.607 &	0.577 &	0.538 &	\textbf{11.37} & \textbf{6.76} &	0.603 &	0.586 &	0.560 &	\textbf{7.13} &	\textbf{4.44} \\
        \hline
        7 &	0.686 &	0.651 &	0.613 &	\textbf{10.64} & \textbf{5.84} &	0.717 &	0.696 &	0.660 &	\textbf{7.95} &	\textbf{5.17} \\
        \hline
        8 &	0.820 &	0.776 &	0.730 &	\textbf{10.98} & \textbf{5.93} &	0.764 &	0.742 &	0.702 &	\textbf{8.12} &	\textbf{5.39} \\
        \hline
        9 &	0.884 &	0.835 &	0.781 &	\textbf{11.65} & \textbf{6.47} &	0.878 &	0.851 &	0.810 &	\textbf{7.74} &	\textbf{4.82} \\
        \hline
        10 &	0.994 &	0.938 &	0.884 &	\textbf{11.07} & \textbf{5.76} &	0.935 &	0.907 &	0.862 &	\textbf{7.81} &	\textbf{4.96} \\
        \hline
        11 &	1.106 &	1.044 &	0.993 &	\textbf{10.22} & \textbf{4.89} &	1.050 &	1.015 &	0.969 &	\textbf{7.71} &	\textbf{4.53} \\
        \hline
        12 &	1.247 &	1.172 &	1.103 &	\textbf{11.55} & \textbf{5.89} &	1.149 &	1.112 &	1.056 &	\textbf{8.09} &	\textbf{5.04} \\
        \hline
        13 &	1.388 &	1.305 &	1.241 &	\textbf{10.59} & \textbf{4.90} &	1.227 &	1.184 &	1.130 &	\textbf{7.91} &	\textbf{4.56} \\
        \hline
        14 &	1.456 &	1.368 &	1.302 &	\textbf{10.58} & \textbf{4.82} &	1.511 &	1.457 &	1.398 &	\textbf{7.48} &	\textbf{4.05} \\
        \hline
        15 &	1.601 &	1.501 &	1.421 &	\textbf{11.24} & \textbf{5.33} &	1.458 &	1.407 &	1.340 &	\textbf{8.09} &	\textbf{4.76} \\
       \hline
        16 &	1.687 &	1.583 &	1.505 &	\textbf{10.79} & \textbf{4.93} &	1.551 &	1.496 &	1.430 &	\textbf{7.80} &	\textbf{4.41} \\
        \hline
        17 &	1.867 &	1.766 &	1.704 &	\textbf{8.73} &	\textbf{3.51}  &	1.645 &	1.586 &	1.521 &	\textbf{7.54} &	\textbf{4.10} \\
        \hline
        18 &	1.741 &	1.631 &	1.544 &	\textbf{11.32} & \textbf{5.33} &	1.739 &	1.679 &	1.614 &	\textbf{7.19} &	\textbf{3.87} \\
        \hline
        19 &	1.834 &	1.720 &	1.635 &	\textbf{10.85} & \textbf{4.94} &	1.836 &	1.773 &	1.710 &	\textbf{6.86} &	\textbf{3.55} \\
        \hline
        20 &	1.928 &	1.811 &	1.731 &	\textbf{10.22} & \textbf{4.42} &	1.933 &	1.870 &	1.808 &	\textbf{6.47} &	\textbf{3.32} \\
        \hline \hline
 \textbf{AVG} &	1.077 &	1.015 &	0.962 &	\textbf{10.62} & \textbf{5.56} &	1.032 &	0.999 &	0.956 &	\textbf{7.21} &	\textbf{4.37} \\
        \hline		
		\end{tabular}
\end{table*}
\begin{table*} []
	\renewcommand{\arraystretch}{1.3}
	\caption{Bitrate reduction ($BR$) achieved by AC-DPCM-SQ for $Goldhill$ and $Peppers$.}
	\label{table:Table3}
	\centering
	\begin{tabular}{|c||c|c|c|c|c||c|c|c|c|c|}
		\hline
		\multirow{2}{*}{\bfseries \shortstack{Subrate\\Index}} & \multicolumn{5}{|c||}{\bfseries \shortstack{$Goldhill$}} &  \multicolumn{5}{|c|}{\bfseries\shortstack{$Peppers$}} \\
        \cline{2-11}
        & \bfseries \shortstack{$Meth1$\\$[bpp]$} & \bfseries \shortstack{$Meth2$\\$[bpp]$} & \bfseries \shortstack{$Meth3$\\$[bpp]$} & \bfseries \shortstack{$BR_{1,3}$\\$[\%]$} & \bfseries \shortstack{$BR_{2,3}$\\$[\%]$} & \bfseries \shortstack{$Meth1$\\$[bpp]$} & \bfseries \shortstack{$Meth2$\\$[bpp]$} & \bfseries \shortstack{$Meth3$\\$[bpp]$} & \bfseries \shortstack{$BR_{1,3}$\\$[\%]$} & \bfseries \shortstack{$BR_{2,3}$\\$[\%]$} \\
        \hline\hline
        1 &	0.152 &	0.153 &	0.152 &	\textbf{0.00} &	\textbf{0.65} &	0.156 &	0.152 &	0.148 &	\textbf{5.13} &	\textbf{2.63} \\
        \hline
        2 &	0.244 &	0.245 &	0.240 &	\textbf{1.64} &	\textbf{2.04} &	0.246 &	0.238 &	0.227 &	\textbf{7.72} &	\textbf{4.62}\\
        \hline
        3 &	0.391 &	0.390 &	0.380 &	\textbf{2.81} &	\textbf{2.56} &	0.381 &	0.369 &	0.348 &	\textbf{8.66} &	\textbf{5.69}\\
        \hline
        4 &	0.417 &	0.413 &	0.402 &	\textbf{3.60} &	\textbf{2.66} &	0.458 &	0.443 &	0.421 &	\textbf{8.08} &	\textbf{4.97}\\
        \hline
        5 &	0.512 &	0.507 &	0.494 &	\textbf{3.52} &	\textbf{2.56} &	0.514 &	0.495 &	0.470 &	\textbf{8.56} &	\textbf{5.05}\\
        \hline
        6 &	0.578 &	0.569 &	0.553 &	\textbf{4.33} &	\textbf{2.81} &	0.638 &	0.612 &	0.580 &	\textbf{9.09} &	\textbf{5.23}\\
        \hline
        7 &	0.659 &	0.650 &	0.634 &	\textbf{3.79} &	\textbf{2.46} &	0.750 &	0.720 &	0.688 &	\textbf{8.27} &	\textbf{4.44}\\
        \hline
        8 &	0.784 &	0.775 &	0.758 &	\textbf{3.32} &	\textbf{2.19} &	0.780 &	0.747 &	0.710 & \textbf{8.97} &	\textbf{4.95}\\
        \hline
        9 &	0.898 &	0.881 &	0.857 &	\textbf{4.57} &	\textbf{2.72} &	0.931 &	0.888 &	0.844 &	\textbf{9.34} &	\textbf{4.95}\\
        \hline
        10 &	0.958 &	0.939 &	0.912 &	\textbf{4.80} &	\textbf{2.88} &	1.053 &	1.006 &	0.961 &	\textbf{8.74} &	\textbf{4.47}\\
        \hline
        11 &	1.022 &	1.003 &	0.974 &	\textbf{4.70} &	\textbf{2.89} &	1.121 &	1.068 &	1.022 &	\textbf{8.83} &	\textbf{4.31}\\
        \hline
        12 &	1.267 &	1.242 &	1.211 &	\textbf{4.42} &	\textbf{2.50} &	1.179 &	1.125 &	1.080 &	\textbf{8.40} &	\textbf{4.00}\\
        \hline
        13 &	1.252 &	1.227 &	1.189 &	\textbf{5.03} &	\textbf{3.10} &	1.271 &	1.207 &	1.148 &	\textbf{9.68} &	\textbf{4.89}\\
        \hline
        14 &	1.559 &	1.529 &	1.493 &	\textbf{4.23} &	\textbf{2.35} &	1.477 &	1.405 &	1.349 &	\textbf{8.67} &	\textbf{3.99}\\
        \hline
        15 &	1.644 &	1.616 &	1.581 &	\textbf{3.83} &	\textbf{2.17} &	1.477 &	1.410 &	1.358 &	\textbf{8.06} &	\textbf{3.69}\\
        \hline
        16 &	1.717 &	1.687 &	1.649 &	\textbf{3.96} &	\textbf{2.25} &	1.563 &	1.485 &	1.419 &	\textbf{9.21} &	\textbf{4.44}\\
        \hline
        17 &	1.793 &	1.762 &	1.722 &	\textbf{3.96} &	\textbf{2.27} &	1.643 &	1.564 &	1.500 &	\textbf{8.70} &	\textbf{4.09}\\
        \hline
        18 &	1.888 &	1.860 &	1.821 &	\textbf{3.55} &	\textbf{2.10} &	1.882 &	1.793 &	1.727 &	\textbf{8.24} &	\textbf{3.68}\\
        \hline
        19 &	1.982 &	1.960 &	1.921 &	\textbf{3.08} &	\textbf{1.99} &	2.061 &	1.990 &	1.937 &	\textbf{6.02} &	\textbf{2.66}\\
        \hline
        20 &	1.872 &	1.839 &	1.797 &	\textbf{4.01} &	\textbf{2.28} &	1.953 &	1.856 &	1.779 &	\textbf{8.91} &	\textbf{4.15}\\
       \hline \hline
       \textbf{AVG} & 1.079 & 1.062 & 1.037 & \textbf{3.66} & \textbf{2.37} & 1.077 & 1.029 & 0.986 & \textbf{8.36}	& \textbf{4.35} \\
      \hline
		\end{tabular}
\end{table*}

\begin{figure}[]
 	\centering
 	\subfloat[$Lenna$]{\label{Fig:LennaRD}
 		\includegraphics[width=0.45\textwidth]{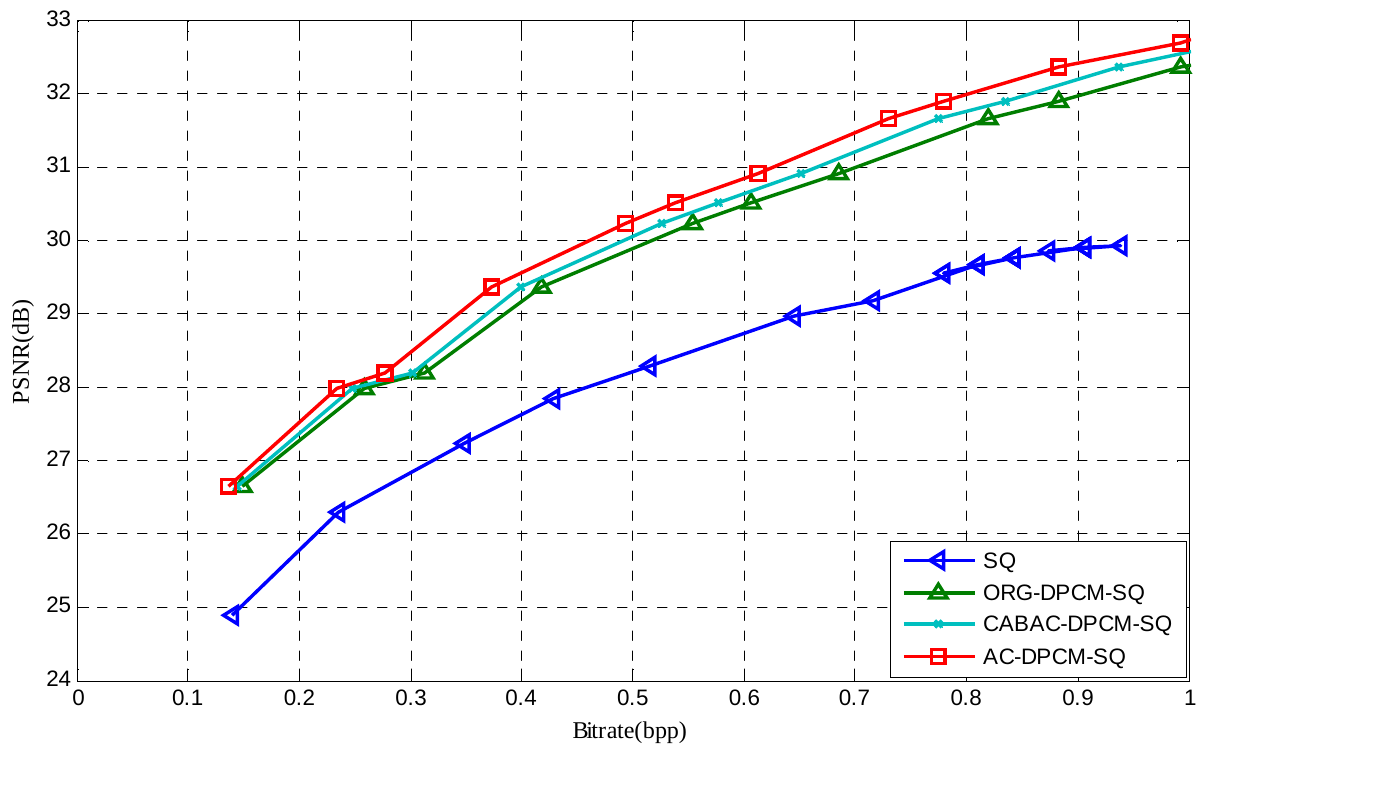}
 	}
 \\
 	\subfloat[$Barbara$]{\label{Fig:BarbaraRD}
 		\includegraphics[width=0.45\textwidth]{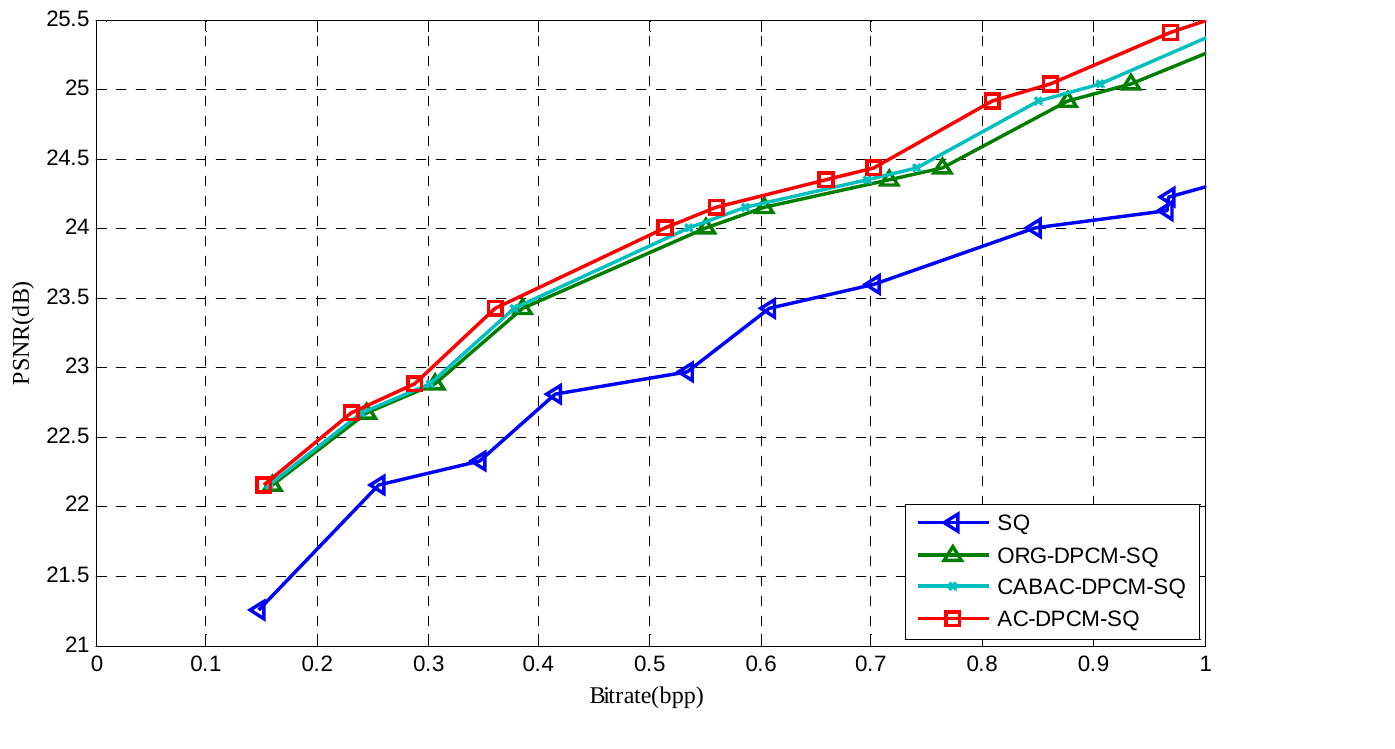}%
 	}
 \\
 	\subfloat[$Goldhill$] {\label{Fig:GoldhillRD}
 		\includegraphics[width=0.45\textwidth]{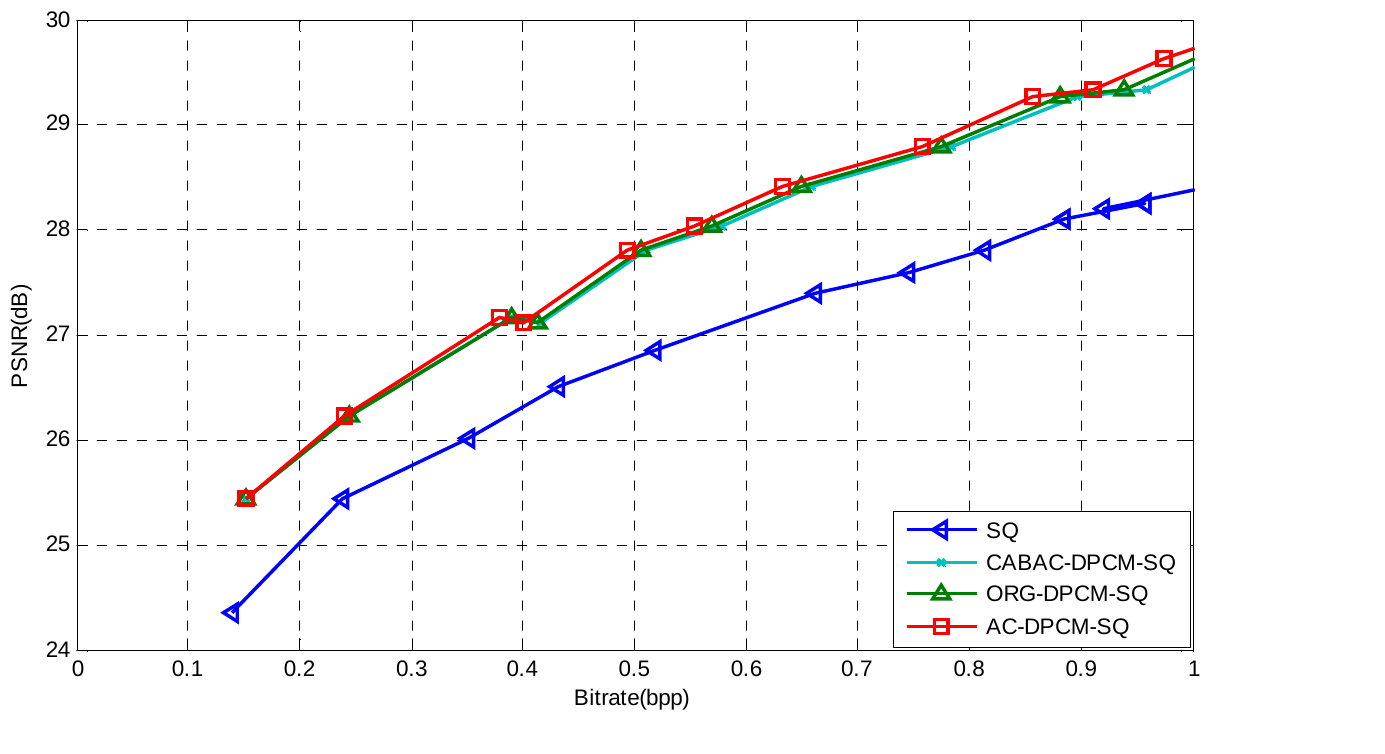}%
 	}
 \\
    \subfloat[$Peppers$] {\label{Fig:PepperRD}
 		\includegraphics[width=0.45\textwidth]{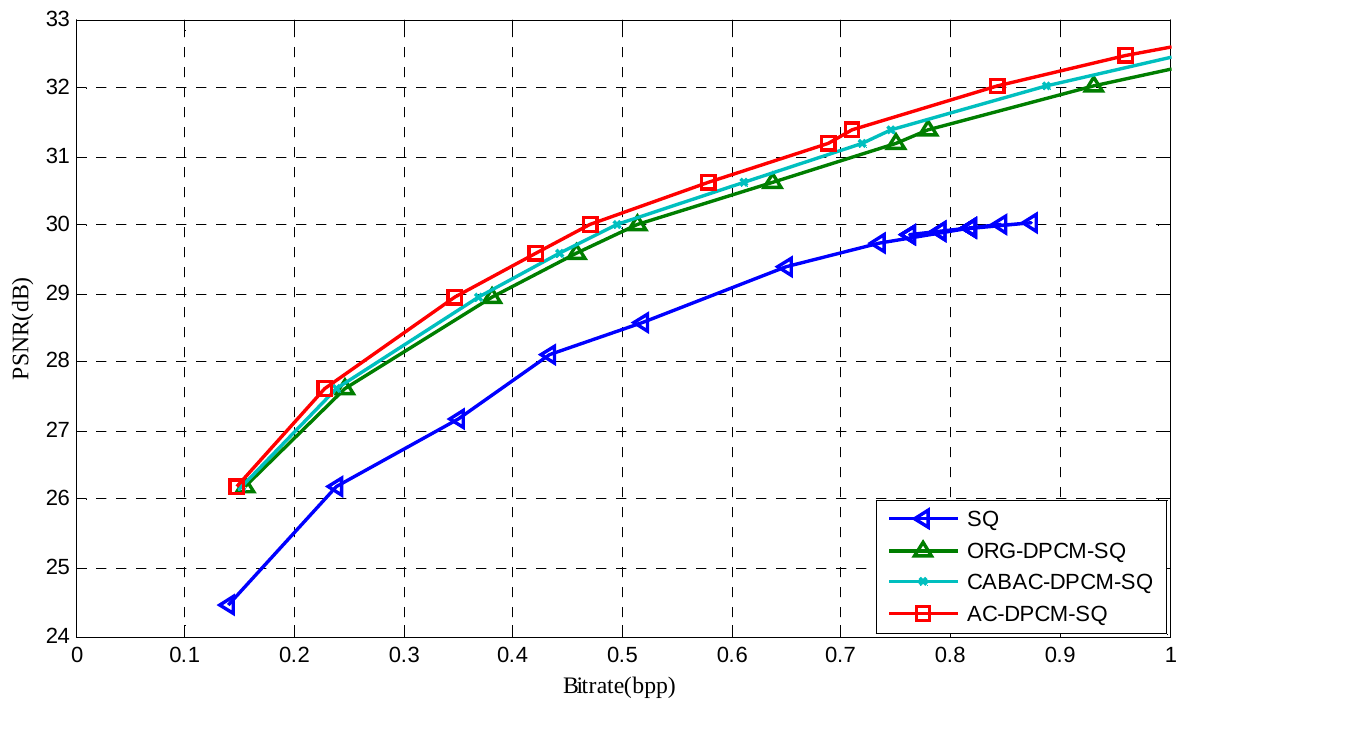}%
 	}
 	\caption{Rate-Distortion (RD) curves for the test images. }
 	\label{Fig:RDCurves}
 \end{figure}

Figs. \ref{Fig:RDCurves} present the rate-distortion (RD) curves of AC-DPCM-SQ for a bitrate ranging from 0.1 $bpp$ to 1.0 $bpp$, in which $SQ$ denotes the method applying uniform scalar quantization along to BCS measurements. In these figures, the bitrates for $SQ$ are also estimated by the zero order entropy of the quantization index. Here, only one image CS recovery algorithm, namely SPL [14] is used to effectuate the CS recovery from the decoded CS measurements, because the arithmetic coding is a lossless coding method. The implementations of the DPCM-plus-SQ and SPL can be found at the BCS-SPL website [15].

\section{Conclusion}
An arithmetic coding scheme is proposed in this paper for the block-based compressive sensing (BCS) of images. According to the statistical information of the quantization index, the proposed arithmetic coding scheme adopts the syntax element \textit{significant\_map} to indicate location of the significant components within a quantization index vector. For each significant component, the syntax elements \textit{abs\_coeff\_level\_minus1} and \textit{sign\_flag} are used to indicate its magnitude and sign, respectively. For \textit{significant\_map} and \textit{abs\_coeff\_level\_minus1}, only one context model is used due to the randomness in the generation of the CS measurements. To efficiently coding these syntax elements, the binary arithmetic coding engine M-coder is used, which can capture the local statistical characteristics of these syntax elements with the probability update module. To the best of our knowledge, this is the first time that the arithmetic coding scheme is designed for the measurement compression within BSC of images. Experimental results demonstrate that further rate-distortion performance can be achieved compared to original DPCM-plus-SQ scheme and transform coefficient coding in CABAC.


%

%
%

\ifCLASSOPTIONcaptionsoff
  \newpage
\fi

\end{document}